\documentclass[letterpaper]{article} 
\usepackage{aaai23}  
\usepackage{times}  
\usepackage{helvet}  
\usepackage{courier}  
\usepackage[hyphens]{url}  
\usepackage{graphicx} 
\usepackage{natbib}  
\usepackage{caption} 
\usepackage{algorithm}
\usepackage{algorithmic}

\usepackage{newfloat}
\usepackage{listings}
\DeclareCaptionStyle{ruled}{labelfont=normalfont,labelsep=colon,strut=off} 
\floatstyle{ruled}
\newfloat{listing}{tb}{lst}{}
\floatname{listing}{Listing}
\title{A Multi-Platform Collection of Social Media Posts \\about the 2022 U.S. Midterm Elections}

\author{
Rachith Aiyappa,$^*$\textsuperscript{\rm 1}
Matthew R. DeVerna,$^*$\textsuperscript{\rm 1}
Manita Pote,$^*$\textsuperscript{\rm 1}
Bao Tran Truong$^*$,\textsuperscript{\rm 1} \\
Wanying Zhao,$^*$\textsuperscript{\rm 2}
David Axelrod,\textsuperscript{\rm 1}
Aria Pessianzadeh,\textsuperscript{\rm 1}
Zoher Kachwala,\textsuperscript{\rm 1}\\
Munjung Kim,\textsuperscript{\rm 2}
Ozgur Can Seckin,\textsuperscript{\rm 1}
Minsuk Kim,\textsuperscript{\rm 2}
Sunny Gandhi,\textsuperscript{\rm 2} \\
Amrutha Manikonda,\textsuperscript{\rm 2}
Francesco Pierri,\textsuperscript{\rm 3}
Filippo Menczer,\textsuperscript{\rm 1} and
Kai-Cheng Yang\textsuperscript{\rm 1}
\\}

\affiliations{
    \textsuperscript{\rm 1}Observatory on Social Media, Indiana University, Bloomington, USA\\
    \textsuperscript{\rm 2}Luddy School of Informatics, Computing, and Engineering, Indiana University, Bloomington, USA\\
    \textsuperscript{\rm 3}Dipartimento di Elettronica, Informazione e Bioingegneria, Politecnico di Milano, Italy\\
    \{racball, mdeverna, potem, baotruon, zhaowany, daaxelro, apessian, zkachwal, munjkim,\\oseckin, mk139, sugandhi, amanikon\}@iu.edu, francesco.pierri@polimi.it, \{fil, yangkc\}@iu.edu\\
}

\newcommand{\hashtag}[1]{{\textit{\##1}}}

\newcommand{\projectnameshort}{\textit{MEIU22}}

\begin{document}

\maketitle

\def\thefootnote{*}\footnotetext{These authors contributed equally to this work; ordered alphabetically}

\def\thefootnote{\arabic{footnote}}

\setcounter{footnote}{0}

\begin{abstract}
Social media are utilized by millions of citizens to discuss important political issues.
Politicians use these platforms to connect with the public and broadcast policy positions.
Therefore, data from social media has enabled many studies of political discussion. 
While most analyses are limited to data from individual platforms, people are embedded in a larger information ecosystem spanning multiple social networks.
Here we describe and provide access to the Indiana University 2022 U.S. Midterms Multi-Platform Social Media Dataset (\projectnameshort{}), a collection of social media posts from Twitter, Facebook, Instagram, Reddit, and 4chan.
\projectnameshort{} links to posts about the midterm elections based on a comprehensive list of keywords and tracks the social media accounts of 1,011 candidates from October 1 to December 25, 2022.
We also publish the source code of our pipeline to enable similar multi-platform research projects.
\end{abstract}

\section{Introduction}

As social interactions increasingly happen online, social media have become important for political discussions.
Most Americans use at least one social media platform throughout the day~\cite{auxier2021pew}, with a sizable proportion of online discussion related to politics~\cite{bestvater2022pew}, especially during election seasons~\cite{bestvater2022pew_fact}.
Government officials heavily leverage social media to interact with their constituencies~\cite{shah2021pew}.
These platforms, therefore, offer a fertile ground for studying public discourse.
For instance, some research mines social media data to learn how misinformation spreads during elections~\cite{shao2018spread,grinberg2019fake} and pandemics \cite{pierri2023one,gallotti2020assessing} and how polarization emerges in digital communities~\cite{waller2021quantifying}.

Most research on social media focuses on data collected from individual platforms---especially Twitter, owing to its openness in the past.
However, this approach can only provide a limited understanding of the larger information ecosystem. As platforms and their user bases differ, results from one social network do not necessarily generalize to others. 
For instance, previous research suggests that polarization processes exhibit different patterns~\cite{yarchi2021political} and that audiences respond differently to similar campaign strategies by the same candidates~\cite{bossetta2022cross} across platforms.
Moreover, moderation efforts by individual platforms might not be sufficient to curb the spread of malicious content given the interconnection among social media~\cite{johnson2019hidden,velasquez2021online}.
Linking data from different social media sites can therefore reveal disinformation campaigns sharing content across platforms~\cite{wilson2020cross,golovchenko2020cross,pierri2022propaganda,infodemic-twitter-vs-facebook}.
As many people use multiple social media~\cite{hardy2018moderating, primack2017use}, they may be even more vulnerable to these cross-platform influence campaigns.

This evidence highlights the need to analyze multiple platforms simultaneously when studying social media.
However, the lack of data hinders such efforts.
In some cases, researchers may have to string together data collected at different times, for different purposes, and using different methods~\cite{lukito2020coordinating}.
Such data may not be comprehensive as it is retrieved retroactively.
Further hurdles to multi-platform data analysis arise from the lack of unified data-sharing protocols~\cite{pasquetto2020tackling}.
We attempt to address these challenges by providing a topic-consistent dataset with broad coverage from multiple platforms during the same time period.

The \projectnameshort{} dataset is a collection of posts from Twitter, Facebook, Instagram, Reddit, and 4chan during the 2022 U.S. midterm election season. We focus on the first four because they are among the top ten most popular platforms used by Americans~\cite{auxier2021pew}. In particular, Twitter and Facebook are increasingly being used for political communication~\cite{stier2018election}. Although not as mainstream as the others, 4chan underlies far-right extremist movements that might affect election results~\cite{baele2020uncovering}. 

To maximize coverage, we first deploy a snowball sampling procedure to identify relevant keywords from multiple platforms in an iterative manner. In addition, we manually compile a list of social media handles for 1,011 candidates with social media presence from any U.S. state.  
Using this information, we build a data collection workflow that fetches posts continuously or through periodic searches, based on the functionalities of the application programming interface (API) endpoints available from different platforms.

The remainder of this paper presents our system architecture, data collection, sources, and processing.
Lastly, we discuss the limitations and potential applications of the data.
For example, our dataset will allow researchers to study public discourse around the 2022 U.S. midterm elections.
The data can also be used to analyze the information diffusion process and potential manipulation on multiple social networks at once.
Furthermore, the social media handles and activities of the candidates included in this dataset allow for in-depth analyses of their public communication strategies.

We provide public access to the dataset as well as the source code of our data collection framework to facilitate replication in other contexts (\url{github.com/osome-iu/MEIU22}).

\section{System architecture}

\begin{figure*}
    \centering
    \includegraphics[width=0.9\textwidth]{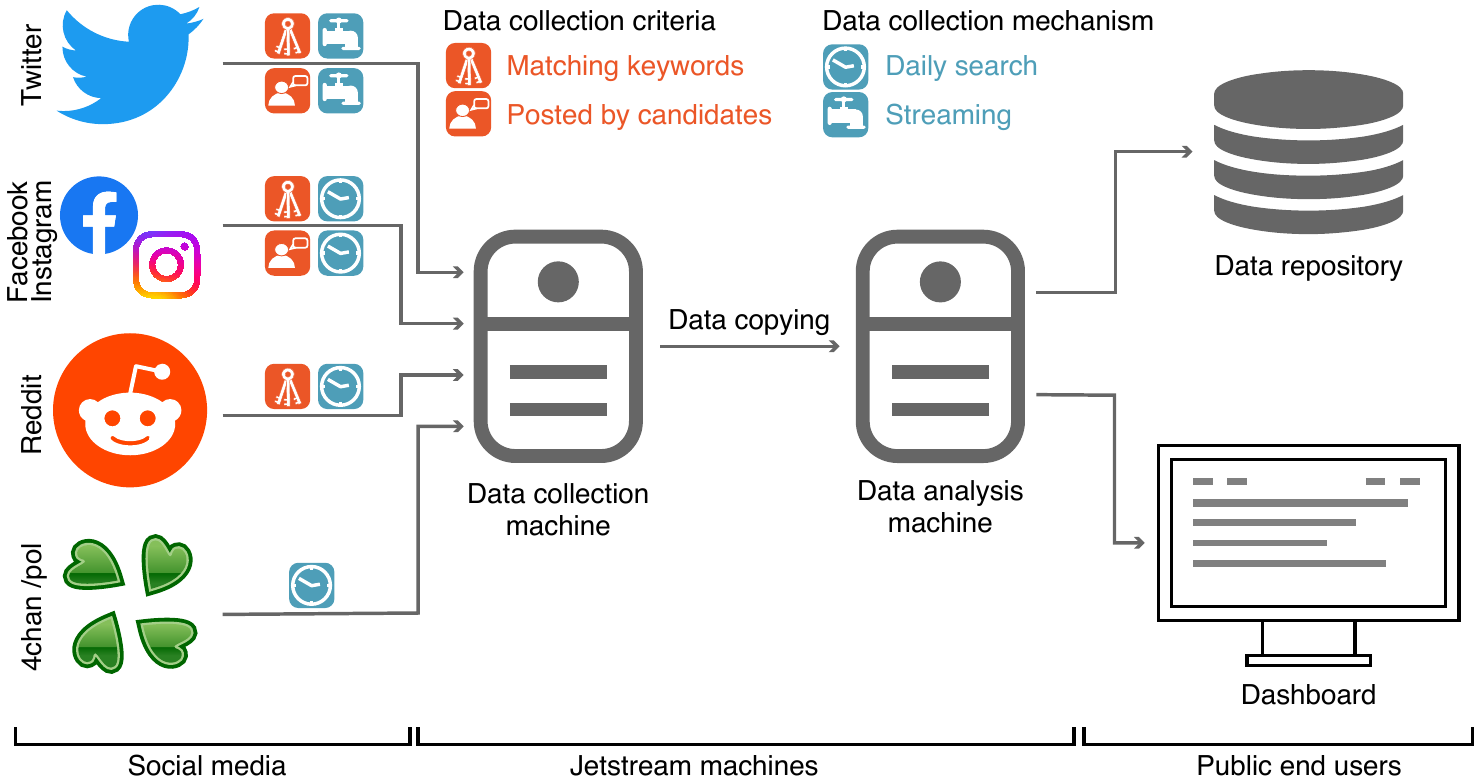}
    \caption{
    Architecture of the \projectnameshort{} data collection and analysis system.
    Data flows in the direction of the arrows.
    }
    \label{fig:ht_serverarchtecture}
\end{figure*}

The architecture of our data collection system is illustrated in Figure~\ref{fig:ht_serverarchtecture}.
This infrastructure is hosted by Extreme Science and Engineering Discovery Environment (XSEDE) Jetstream virtual machines (VMs)~\cite{xsede,stewartJetstreamSelfprovisionedScalable2015}, which can be replaced with any server that has access to the internet and enough storage space.

We distribute the tasks to two VMs.
The data \textit{collection machine} is responsible for collecting the raw data and copying it to the \textit{analysis machine} periodically.
The collected data is also backed up to Indiana University's Scholarly Data Archive, a fast tape storage system, to ensure robustness (not illustrated).
Depending on the functionalities of the API endpoints, we use different strategies to collect data from different platforms, as explained in detail in the following section.
Computationally expensive tasks such as data cleaning and analysis are executed on the \textit{analysis machine}.
This machine is also responsible for hosting a dashboard\footnote{\url{osome.iu.edu/tools/midterm22}} to share insights obtained from the data with the public.

\section{Data collection}

Our dataset consists of two parts: (1) general social media posts discussing election-related issues, and (2) posts published by U.S. congressional candidates.
For the first part, we employ a keyword-based data collection approach, using the same keyword list for different platforms to ensure that the data is comparable.
For the second part, we compile a list of the social media handles of all the candidates and use it to track their social media activity.
In the following, we describe the procedure we adopt to obtain relevant keywords and the midterm candidate list.
Additionally, we provide details on how the data is collected on each social media platform.

\subsection{Keyword list}

\begin{table}
\centering
\begin{tabular}{l|p{5cm}}
    \hline
    Date of inclusion & Keywords \\
    \hline
    2022-09-16* & midterm~election, 2022~midterm, 2022~election, midterm~2022 \\
    \hline
    2022-09-20* & vote~2022, vote~midterm, vote~november, midterm~november, vote~republicans, vote~democrat,\\ & voteblue, votered \\
    \hline
    2022-09-30* & november~republicans, november democrats, absentee~vote,\\& absentee~ballot, mail~in~vote, mail~in~ballot \\
    \hline
    2022-10-07 & mail~ballot, october~surprise \\
    \hline
    2022-11-04 & ivoted, red wave, blue wave \\
    \hline
\end{tabular}%
\caption{
    List of keywords used to collect data from different platforms.
    Only the basic form of each keyword is shown here.
    For the full list containing all the variants, please refer to the GitHub repository.
    Asterisks indicate the initial phase.
    The term ``ivoted'' is removed on November 11, our only keyword removal.
}
\label{table:keyword_list}
\end{table}

We build the keyword list during an initial curation phase between September 16 and September 30, 2022.
We employ a snowball sampling procedure~\cite{covaxxy_deverna_2021,di2022vaccineu}, as detailed below.
This leads to the collection of phrases appearing most often in discussions about the midterm elections.
Such a collection is performed separately on each platform (Twitter, Facebook, Instagram, and Reddit), but all the phrases are merged into a single list.

We perform three rounds of the snowball procedure to iteratively build the list of keywords/phrases (see Table~\ref{table:keyword_list}).
We start from a short list containing a number of unambiguous \textit{seed} phrases related to the midterm elections. 
These seeds are used as queries for the APIs of different platforms.
Each snowball round consists of collecting data for multiple days and then identifying the top 50 unigrams and top 50 bigrams (not necessarily consecutively) that co-occur with any of the phrases in the current list, for each platform.

Before being added to the list, potential phrases are manually reviewed by three of the authors (R.A., M.R.D., and K.-C.Y.) for inclusion in the list for the next round.
They are included only after considering relevance and precision with respect to capturing discussion related to the U.S. midterm elections.
For example, it is common for issue-related phrases to appear within the top-ranked uni/bigrams.
Some of these phrases, e.g., ``abortion,'' or ``abortion ban,'' capture general discussions about abortion that are not necessarily connected to the elections, and are therefore excluded.
Other issue-related phrases, such as ``mail-in vote,'' lead to very few false positives, and are therefore included.
Note that due to the nature of the streaming API, newly added keywords only affect subsequent matching.

Once a phrase enters the keyword list, different variants of it are also added to ensure complete coverage of the semantic meaning of this phrase and because different platforms differ in the way their APIs match keywords (see individual platform-related sections for details). 
For example, after identifying the phrase ``november midterm,'' we also include ``midterm november.''
Similarly, we add plurals of the phrases when they have a proper semantic meaning (e.g., ``mail-in vote'' is complemented with ``mail-in votes'').

Data collection from different platforms occurs between October 1 and December 25 using the keyword list.
During this period, we repeat the snowball sampling procedure every seven days but with stricter criteria for adding new phrases to keep the list stable.
For example, new phrases are only included if they capture stand-out ``viral'' events that dominate political discussion at the time.

\subsection{Candidate list}

We obtain the social media handles of the U.S. Senate and House election candidates from ballotpedia.org.\footnote{\url{ballotpedia.org/List_of_congressional_candidates_in_the_2022_elections}}
For each candidate, we collect the following information: name, party affiliation, election type (Senate or House), house candidate district, Twitter handle, Facebook and/or Instagram pages, YouTube channel, and links to the candidate's campaign/official websites.
The personal pages of the candidates are not added to the list as we assume they do not contain election-related content.
Since Twitter users are allowed to change their usernames, we also extract the unique numerical ID for each handle.
We exclude candidates who have already lost in their primary elections, keeping information for those that have advanced to the general elections in November 2022.
In total, the list contains information about 4,508 social platform handles from as many as 1,011 candidates running for the 2022 U.S. House and Senate elections.
The list is also shared on the GitHub repository.

\subsection{Twitter}

Twitter data is collected using the \texttt{tweepy} Python library, which uses the Twitter V1 filter streaming API endpoint.\footnote{\url{developer.twitter.com/en/docs/twitter-api/v1/tweets/filter-realtime/overview}}
This endpoint allows us to collect all public tweets containing our keywords.
In this process, the texts of the tweets and certain entity fields are considered for matches.
These entity fields include hashtags, the expanded and display URLs, and the screen name for user mentions.\footnote{\url{developer.twitter.com/en/docs/twitter-api/v1/tweets/filter-realtime/guides/basic-stream-parameters}}

We also use the filter streaming API endpoint to collect tweets from the candidates.
This collection process did not start until October 24, 2022, so we additionally fetch all posts by these candidates since June 1, 2022 with the user timeline endpoint.\footnote{\url{developer.twitter.com/en/docs/twitter-api/v1/tweets/timelines/api-reference/get-statuses-user_timeline}}

To abide by Twitter's terms of service, we are only allowed to publicly share the tweet IDs of the retrieved tweets.
The dataset can be re-hydrated by querying the Twitter API directly or using tools like Hydrator\footnote{\url{github.com/DocNow/hydrator}} or twarc.\footnote{\url{github.com/DocNow/twarc}}
Although we use the Twitter V1 API to collect the data, one can alternatively employ the V2 API\footnote{developer.twitter.com/en/docs/twitter-api} to re-hydrate the data.

\subsection{Facebook and Instagram}

Facebook and Instagram data is collected from CrowdTangle, a public insights tool owned and operated by Facebook~\cite{crowdtangle}.
Specifically, we query the \texttt{/posts/search} endpoint,\footnote{\url{github.com/CrowdTangle/API/wiki/Search}} retrieving posts from \textit{both} Facebook and Instagram simultaneously.
These searches are not case-sensitive.
As this platform does not provide access to real-time data, we retrieve posts for the previous day every morning (all dates based on Coordinated Universal Time, or UTC).

Facebook posts from candidates are collected using the \texttt{/posts} endpoint\footnote{\url{github.com/CrowdTangle/API/wiki/Post}} that downloads all public candidate pages and groups in a CrowdTangle list. This list, which we manually curate, manages access to all posts from pages and groups with a web-based interface.
We again adopt the practice of downloading all posts from the previous day, each morning beginning on October 31.
Earlier data since June 6, 2022 (inclusive) was fetched between October 24 and October 29.

Note that CrowdTangle only tracks data from \textit{public} Facebook and Instagram pages and groups, so we do not have complete visibility into the platform activity.
This also means that we are unable to track some candidate accounts not indexed by CrowdTangle.
We are only allowed to share the URL of collected posts, which can be used to access the data publicly.

\subsection{Facebook and Instagram advertisements}

We collect information about advertisements on Facebook and Instagram using the Meta Ad library API.\footnote{\url{facebook.com/ads/library/api}}
This API provides a single endpoint \texttt{ads\_archive} to search all ads stored in the Ad library.
Queries are made using the same keyword list described above.
The Meta Ad library captures text in various data fields, such as text, image, audio, video, and the ``call-to-action,'' of an advertisement.
Thus, if any of an advertisement's fields contain any of the matching phrases, data for that advertisement is captured within our dataset.
Each day, we collect all the advertisements that are labeled as political or issue-related, and that are delivered in the U.S.
Meta implemented a restriction on election-related advertisements between November 1 and 8,\footnote{\url{developers.facebook.com/blog/post/2022/09/28/upcoming-restriction-period-for-us-ads}} leading to a drastic decrease in the data volume (see Figure~\ref{fig:volume}).

According to platform policy, we are allowed to share the raw advertisement data only with researchers or journalists who have a Meta developer account and agree to Meta's platform policy.
Here, we share the list of page IDs in which advertisements are displayed. Users can retrieve the advertisements by querying the API with these IDs.

\subsection{Reddit}

Reddit submissions and comments are collected using the Pushshift API~\cite{baumgartner2020pushshift}.
The API provides two endpoints for retrieving submissions (\texttt{/reddit/search/submission}) and comments (\texttt{/reddit/search/comment}), respectively.\footnote{\url{github.com/pushshift/api}}
We search for all posts that match at least one keyword.
Given that Pushshift only provides historical data, we apply the same strategy as for CrowdTangle to retrieve posts from Reddit.
There are no sharing restrictions for Reddit data, which is publicly available, and we provide access to the entire collection of submissions and comments.

\subsection{4chan}

4chan is an image-based bulletin board where users can create a thread by posting an image and a message to a board and others can reply to it.
For our data collection, we focus on the ``Politically Incorrect'' board \texttt{/pol}~\cite{papasavva2020raiders}.
One of 4chan's features is the ephemerality of the content.
Specifically, threads that receive recent replies are bumped to the top of the board, pushing older threads down.
The \texttt{/pol} board has limited space (21 pages), and once a thread is pushed out of the board, it enters the archive.
Archived threads are static, and can no longer be replied to.
They are deleted after a certain amount of time that depends on the speed at which new threads are archived.

We adopt two approaches to collect the data from 4chan's \texttt{/pol} board.
In the first case, we hope to understand what people are discussing on this board.
So we simply retrieve all the threads from the catalog endpoint\footnote{\url{a.4cdn.org/pol/catalog.json}} every five minutes, starting on October 1, 2022.
Based on some preliminary analysis, it takes at least 15 minutes for a thread to be archived, therefore we should have obtained all the original posts this way.
But one problem with this approach is that we might miss some replies.
For each original post obtained from the catalog endpoint, its five most recent replies are attached as well.
However, it is very common for people to reply to this thread more than five times since the last snapshot, so some replies are missed.

Therefore we deploy a second approach to obtain the full threads, including all replies starting October 11, 2022.
Here, we leverage the archive endpoint\footnote{\url{a.4cdn.org/pol/archive.json}} and the thread endpoint.\footnote{\url{a.4cdn.org/pol/thread}}
The archive endpoint returns a list of threads in the archive at the moment.
By comparing the snapshot of the archive at the current moment with the snapshot at a previous moment (we use 10 minutes in our collection), we can find the threads that have been archived recently.
Next, we query the thread endpoint to fetch the whole tree, i.e., the original post and all its replies. 
Since the archived threads can no longer be updated, the result is final.
We make the data collected with both methods available for the public to use.

\section{Data processing}
\subsection{Cleaning}

We notice that the posts obtained through the keyword-matching approach contain a lot of false positives.
The issue is mainly due to various award events involving voting (e.g., the American Music Awards) and elections in other countries (e.g., the 2022 Gujarat Legislative Assembly election in India).
Twitter suffers more from the first issue, whereas the second one is more pronounced on Meta and Reddit.
Many posts pertaining to these events use the keywords ``vote,'' ``election,'' and ``2022,'' and thus end up in our collection.
On some days, the false positives make up over 50\% of the posts in the raw collection and we accordingly apply extra data-cleaning procedures.

By analyzing the false positives on Twitter, we find that these tweets often contain hashtags referring to specific events.
Take the American Music Awards as an example: most tweets use the hashtag \hashtag{AMAs} together with the names of some artists, such as \hashtag{BTS}.
Therefore, we curate a negative list of hashtags that refer to events irrelevant to the U.S. midterm elections.
For tweets from a given day, we extract the 50 most popular hashtags and manually inspect them to identify the irrelevant ones.
We then add these hashtags to our negative list and exclude the tweets containing any of the hashtags in it.
The procedure is repeated until all 50 most popular hashtags are related to the midterm elections.
We repeat this procedure for all days from October 1 to December 25, 2022.
In the end, we use the full negative list to filter tweets in the entire collection.

Unlike Twitter, posts from Facebook, Instagram, and Reddit tend to contain fewer hashtags.
Therefore, we extract the unigrams (stop words excluded) from the posts for each day, and rank them by their TF-IDF (term frequency–inverse document frequency).
We manually inspect the top 100 to identify irrelevant ones and add them to the negative list.
Finally, we use the resulting negative list to filter the posts in the entire collection.

\subsection{Quality evaluation}

\begin{table}
\centering
\begin{tabular}{lrrr}
\hline
Platform & Overall & Before Nov. 8 & After Nov. 8 \\
\hline
Twitter & 0.89 & 0.94 & 0.85 \\
Meta & 0.88 & 0.96 & 0.82 \\
Reddit & 0.92 & 0.93 & 0.92 \\
\hline
\end{tabular}%
\caption{
Data quality evaluation results.
We report the precision for the whole time period (October 1 -- December 25), before election day (October 1 -- November 8), and after election day (November 8 -- December 25).
}
\label{table:relevance_eval}
\end{table}

To evaluate the quality of the data collected using the keyword-matching approach after cleaning, we perform manual annotation of sampled data.
For Twitter, Meta (we mix the Facebook and Instagram posts together since they are obtained through the same API endpoint), and Reddit (we combine the submissions and comments together), we sampled ten posts from each day's collection between October 1 to December 25 (i.e., 860 posts in total for each platform).
The authors then manually label each post as a true positive or false positive. 
The true positives include posts clearly discussing the U.S. midterm elections and those referring to other events but using midterm-related keywords to gain attention.
The false positives mainly consist of the posts dedicated to the elections and politics in other countries and those voting artists, movies, or games for awards.

With the annotations, we are able to calculate the precision (the number of false positives divided by the total number of posts) for each platform.
The results can be found in Table~\ref{table:relevance_eval}.
We also report the precision before and after the election day as references.
The results suggest that there is more noise in the post-election period.

\subsection{Data volume}

In this section, we briefly characterize the data collected. Table~\ref{tab:stats} summarizes the total number of midterm-related posts from each platform and the number of candidate handles for Twitter and Facebook.

\begin{table}
\centering
\begin{tabular}{lrrr}
\hline
Platform & \# posts (key.) & \# posts (cand.) & \# handles \\
\hline
Twitter & 6,242,412 & 638,448 & 1,237 \\
Facebook & 304,106 & 259,385 & 1,209 \\
Instagram & 35,314 & N/A & N/A\\
Reddit & 160,773 & N/A & N/A\\
Meta Ads & 5,352 & N/A & N/A\\
\hline
\end{tabular}
\caption{
Summary statistics.
We report the number of posts collected using the keyword list, the number of posts from candidates, and the number of candidate handles on different platforms.
Note that some candidates have multiple social media handles on a single platform.
}
\label{tab:stats}
\end{table}

\begin{figure}
    \centering
    \includegraphics[width=\columnwidth]{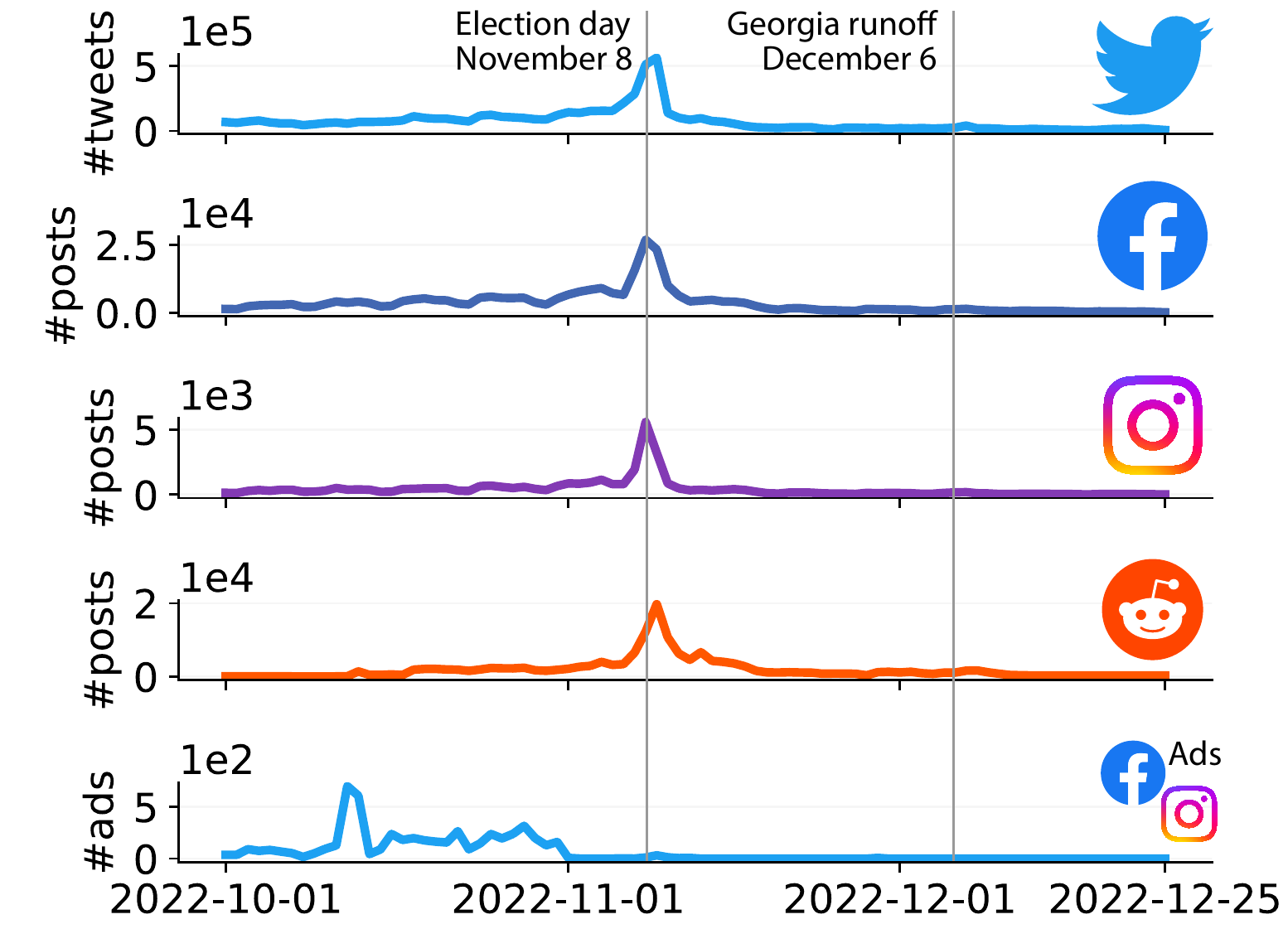}
    \caption{
    Daily volume of midterm-related posts collected through the keyword-matching approach from each platform.
    For Reddit, we combine the number of submissions and comments together.
    For the advertisements, we combine the number on Facebook and Instagram.
    We annotate the election day, i.e., November 8, and the day of the Georgia runoff, i.e., December 6.
    }
    \label{fig:volume}
\end{figure}

Figure~\ref{fig:volume} shows the daily volume of posts collected by matching keywords on different platforms.
We observe volumes ranging from a peak of around 500 thousand posts per day on Twitter to a few thousand on Instagram (recall that CrowdTangle only provides limited data about public Instagram accounts). Despite the difference in volume, these platforms share similar temporal patterns, with the highest number observed around election day.
The case for advertisements on Facebook and Instagram is different, as there are barely any election-related ones after November 1 due to the aforementioned platform policy.

\begin{figure}[ht]
    \centering
    \includegraphics[width=\columnwidth]{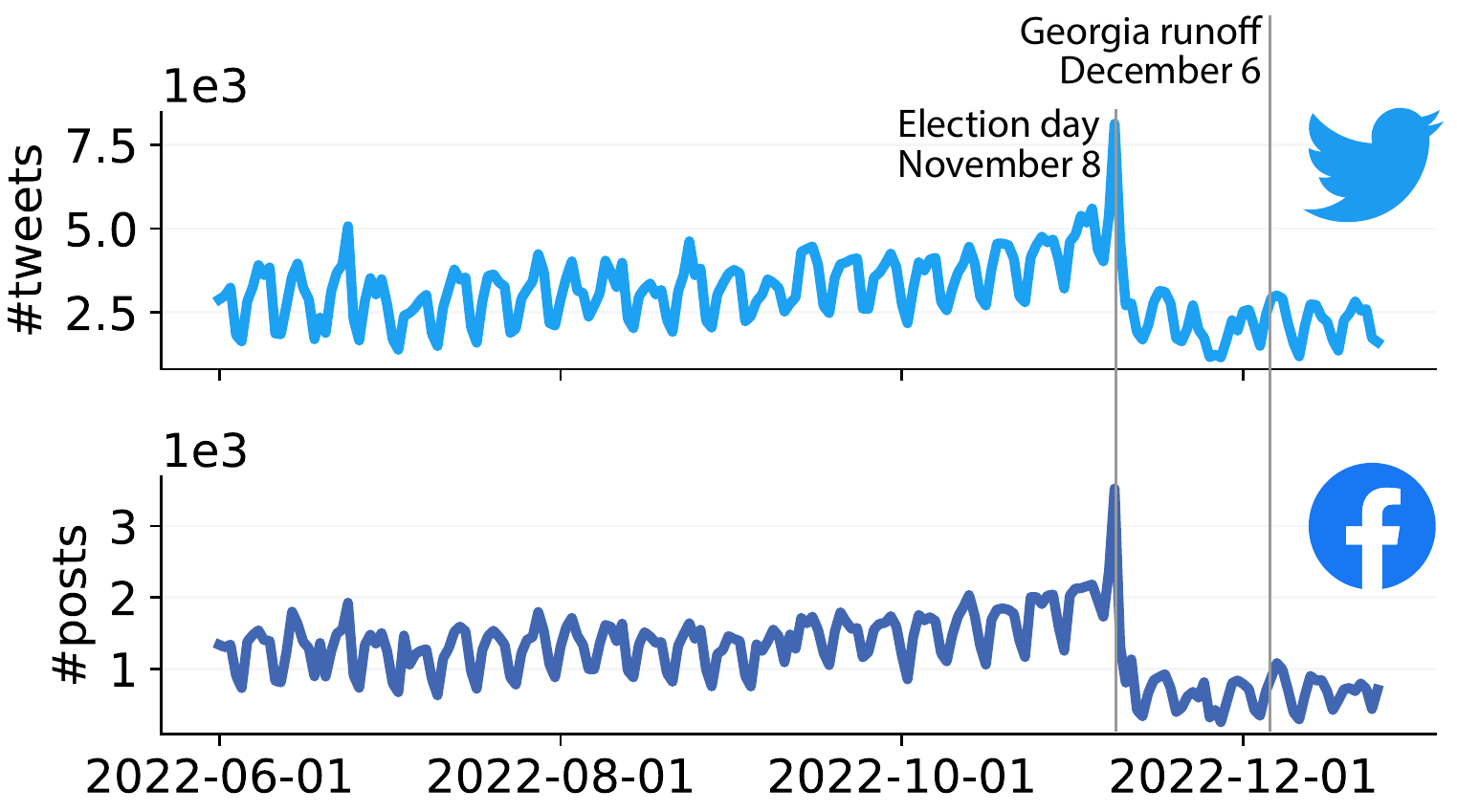}
    \caption{
    Daily volume of tweets and posts generated by the congressional candidates on Twitter and Facebook.
    }
    \label{fig:volume_candidate}
\end{figure}

Figure~\ref{fig:volume_candidate} shows the volume of congressional candidates' tweets and Facebook posts, respectively.
With retrospective search, the data collection covers the time period from June 1 to December 25, 2022.
The time series of data from the two platforms share very similar temporal patterns.

\section{Discussion}

\subsection{Limitations}

Despite our efforts to cover as many social media platforms as possible, our dataset does not include platforms with substantial user bases that might play important roles in the current information ecosystem, such as YouTube and TikTok~\cite{auxier2021pew}.
Emerging platforms such as Parler and Truth Social, which are alternative destinations when users are banned from major platforms, are also not covered ~\cite{stocking2022pew}.
This is mainly because these platforms offer no programmatic methods with which we can obtain data.
Even for some of the data we are able to obtain, we are not allowed to share the content directly.
To retrieve the information, users will need to go through review processes on different social media platforms to obtain access first.
This still poses an obstacle for efforts to study discussions on social media and impedes the progress of open science. 
A case in point is the recent decision by Elon Musk to discontinue free access to Twitter data for research.
As a result, researchers may no longer be able to re-hydrate tweets from the IDs in our dataset.
In this case, we will make the raw data available to researchers upon reasonable request.

In the future, we hope to expand our framework to support more social media platforms.
For instance, TikTok has recently started new programs to increase academic access to the platform.\footnote{\url{newsroom.tiktok.com/en-us/strengthening-our-commitment-to-transparency}}
We also need to adapt the system to the changing API specifications.
For example, Twitter retires its V1 API endpoints on November 23, 2021, but migration to the V2 API endpoints is not trivial.

Another known issue with our dataset is noise stemming from the keyword selection process~\cite{king2017computer}.
Keyword-matching approaches inevitably introduce false positives since some keywords capture both relevant and irrelevant posts.  
Even after the cleaning procedures, we still observe a considerable number of irrelevant posts. In addition, our snowball-sampling procedure might miss some relevant keywords or exclude keywords leading to many irrelevant posts. As a result, the list is comprehensive but not necessarily exhaustive. 

Because of the library used to query the Twitter filter streaming API endpoint, a tweet is matched if all of the terms in any keyword phrase are present in the tweet, regardless of order and case.
For example, for the phrase ``midterms 2022,'' tweet objects containing both ``midterms'' \textit{and} ``2022,'' not necessarily consecutively or in that order, are returned.
This might lead to a higher number of both false and true positives compared to other platforms, creating a potential inconsistency in coverage.
An alternative approach would be to reformulate queries on Twitter to retrieve only tweets where the phrase terms appear consecutively and in the right order.

These sources of noise could affect downstream analyses.
We recommend researchers focus more on the pre-election period when the data quality is higher.
Our dataset includes both the raw data collection and the cleaned version so that researchers can deploy their own data-cleaning processes using more advanced methods, such as natural language processing techniques if needed.

\subsection{Related Datasets}

There are a number of datasets in the literature related to the context of U.S. elections.
Here we briefly describe them.

\textbf{2016 U.S. presidential election:}
Shao et al. provide access to the IDs of over 29M tweets linking to low-credibility and fact-checking websites shared in 2016 and 2017~\cite{hui2018hoaxy,shao2018anatomy}.
They also open source the data collection platform, i.e., Hoaxy.\footnote{\url{hoaxy.osome.iu.edu}}
Similarly, Bovet et al. collect over 250M tweets by querying Twitter's streaming API over a period of 6 months (from June 1 to November 8, 2016), using a list of keywords regarding Donald Trump and Hillary Clinton~\cite{bovet2019influence}.

\textbf{2018 U.S. midterm elections:}
Deb et al. collect two tweet datasets through Twitter's API.
One dataset contains over 250k tweets sharing the hashtag \hashtag{ivoted} posted on November 6, 2018, the election day.
The other one includes over 2M tweets containing election-related hashtags over a period of six weeks~\cite{deb2019perils}.
Similarly, Yang et al. create a dataset of over 60M tweets using a list of 143 relevant hashtags constructed through a snowball sampling approach~\cite{yang2022twitter}.

\textbf{2020 U.S. presidential election:}
Chen et al. provide a continuous collection of over 1B tweets starting May 2019 using Twitter’s streaming API.
They include discussions related to presidential candidates and tweets containing keywords and hashtags in a manually-complied list~\cite{chen2022election2020}.
Abilov et al. focus on election fraud claims and curate a dataset containing over 30M tweets and retweets matching a manually created set of keywords, along with links and metadata of YouTube videos and information of images shared in the tweets~\cite{abilov2021voterfraud2020}.
Kennedy et al., on the other hand, leverage real-time reports of over 400 distinct misinformation stories and use keyword-based searches to collect almost 50M related tweets~\cite{kennedy2022repeat}.

Compared with those datasets, our collection is unique in that it covers the 2022 U.S. midterm elections and spans multiple social media platforms.
The data allows researchers from different disciplines to better understand the online discourse in the most recent U.S. election cycle.
The multi-platform nature of the dataset offers new opportunities to study the whole information ecosystem.
In addition to the dataset, we also make available the source code of our collection framework, which can be used in different contexts.

\subsection{Potential Applications}

Our dataset provides a fertile ground for many studies that might focus on a specific platform or consider multiple platforms simultaneously.
For instance, researchers can compare the spreading patterns of mis/disinformation across different platforms, as has been done comparing Facebook and Twitter~\cite{infodemic-twitter-vs-facebook}, and analyze how malicious content migrates from one social network to another.
The analyses will also shed light on the role of ``superspreaders'' of misinformation, who are often active on multiple platforms simultaneously~\cite{pierri2023one,deverna2022identification}
For those working on detecting and characterizing inauthentic coordinated behaviors, our dataset provides a new test bed for them to implement their methods. 
More importantly, researchers can investigate the presence and the impact on influence operations taking place across different platforms.

By combining the general discussion and the posts from the candidates, researchers can better understand the online interactions between these candidates and their constituencies.
The advertisement data can be used to study politicians' campaigns~\cite{islam2023electionfbAd,pierri2022political}. 
Alternatively, the focus could be the political communication strategies put in place by different candidates on different platforms.
One can also investigate the correlation between online speech and electoral outcomes. 

\bibliography{Aiyappa}

\end{document}